\documentclass[12pt]{article}
\usepackage{epsfig}

\def\red{
}
\def\black{
}

\def\URLtilde{\lower0.2em\hbox{$\tilde{\phantom{a}}$}}
\def\mycomm#1{\hfill\break\strut\kern-3em{\red\tt ====> #1\black}\hfill\break}
\def\mycommNL#1{\strut\kern-3em{\red\tt ====> #1\black}\hfill\break}
\def\ds{\displaystyle}

\catcode`\@=11 % This allows us to modify PLAIN macros.
\def\lsim{\mathrel{\mathpalette\@versim<}}
\def\gsim{\mathrel{\mathpalette\@versim>}}
\def\@versim#1#2{\vcenter{\offinterlineskip
        \ialign{$\m@th#1\hfil##\hfil$\crcr#2\crcr\sim\crcr } }}
\catcode`\@=12 % at signs are no longer letters

\def\sbar{\hbox{$\bar s$}}

\def\thetap{\hbox{$\Theta^+$}}
\def\eqref#1{(\ref{#1})}

\def\gray{\special{ps: 0.40 setgray}}
\def\black{\special{ps: 0.0 setgray}}

\newcommand{\mydraft}{
\newcount\timecount
\newcount\hours \newcount\minutes  \newcount\temp \newcount\pmhours

\hours = \time
\divide\hours by 60
\temp = \hours
\multiply\temp by 60
\minutes = \time
\advance\minutes by -\temp
\def\hour{\the\hours}
\def\minute{\ifnum\minutes<10 0\the\minutes
    \else\the\minutes\fi}
\def\clock{
\ifnum\hours=0 12:\minute\ AM
\else\ifnum\hours<12 \hour:\minute\ AM
\else\ifnum\hours=12 12:\minute\ PM
    \else\ifnum\hours>12
     \pmhours=\hours
     \advance\pmhours by -12
     \the\pmhours:\minute\ PM
     \fi
    \fi
\fi
\fi
}
\def\fullclock{\hour:\minute}
\begin{centering}
\gray
\font\Hugett  =cmtt12 scaled\magstep4
\hbox{\Hugett Draft:\today,\clock}
\black
\end{centering}
\vskip -1.7cm
$\phantom{a}$
} % end of \draft definition
\def\beq#1{\begin{equation} \label{#1}}
\def\eeq{\end{equation}}
\def\bra#1{\left\langle #1\right\vert}
\def\ket#1{\left\vert #1\right\rangle}

\usepackage[perpage,symbol*]{footmisc}

\begin{document}
\begin{flushright}
%\strut\vskip-4.0cm
Cavendish-HEP-05/09\\
TAUP 2807/05\\
%WIS/????\\
ANL-HEP-PR-05-47
\end{flushright}
\vskip0.8cm
\begin{center}
{\Large\bf Isospin Analysis of \thetap\ production forbids
$\gamma p \rightarrow \thetap\,K_s$ 
\\ \vrule height 2.5ex depth 0ex width 0pt
and allows $\gamma n \rightarrow
\thetap\,K^-$}
%\mydraft
\vskip0.8cm
{\bf Marek Karliner\,$^{a,b}$\footnote{e-mail: \tt marek@proton.tau.ac.il}\\
\vspace{2mm}
and\\
\vspace*{2mm}
Harry J. Lipkin\,$^{b,c}$\footnote{e-mail: \tt ftlipkin@weizmann.ac.il} }\\
\vskip0.8cm
{\it
$^a\;$Cavendish Laboratory\\
University of Cambridge, UK\\
%\mbox{}\\
and\\
$^b\;$School of Physics and Astronomy \\
Raymond and Beverly Sackler Faculty of Exact Sciences \\
Tel Aviv University, Tel Aviv, Israel\\
\vbox{\vskip 0.0truecm}
$^c\;$Department of Particle Physics \\
Weizmann Institute of Science, Rehovot 76100, Israel \\
and\\
High Energy Physics Division, Argonne National Laboratory \\
Argonne, IL 60439-4815, USA
} % end of \it
\end{center}

\vspace*{0.8cm}
\centerline{\bf Abstract}
\vspace*{4mm}
The discrepancy between \thetap\ photoproduction 
on proton vs. deuteron can be resolved if the photon couples 
much more strongly to $K^+K^-$ than to $K^o \bar K^o$ as indicated by the
experimentally observed asymmetry between $\gamma p \rightarrow \Lambda(1520) K^+$
and  $\gamma  n \rightarrow\Lambda(1520) K_s$.
Significant signal-background interference effects can occur in 
experiments like $\gamma N \rightarrow \bar K \Theta^+$  
which search for the \thetap\ as a narrow $I=0$
resonance in a definite final state  against  a nonresonant
background, with an experimental resolution coarser than the expected
resonance
width. 
We show that when the signal and background have roughly the same
magnitude, destructive interference can easily combine
with a limited experimental resolution to completely destroy
the resonance signal. Whether or not this actually occurs depends
critically on the yet unknown relative phase of the $I=0$ and $I=1$
amplitudes. We discuss the implications for some specific experiments.

\vfill\eject

\section{Introduction - Effects of Possible Neutral Kaon 
\\
Suppression}

The recent experimental discovery \cite{Nakano:2003qx} of an exotic 5-quark $K
N$ resonance \thetap\ with positive strangeness, a mass of $\sim$1540 MeV, a
very small width $\lsim 20$ MeV and a presumed quark configuration $uudd\sbar$ 
has given rise to a number of further experiments \cite{hicks} and a new
interest in theoretical models \cite{jenmalt} for exotic hadrons including
models with diquark structures \cite{NewPenta}. But the controversy between
experimental evidence for and against the existence \cite{cryptopen,ichepproc}
of the \thetap\ remains unresolved. There are also questions about isospin
asymmetry \cite{Hosaka}.

\subsection {\thetap\ photoproduction on proton vs neutron targets}
One important issue to be explained is the
difference and apparent isospin asymmetry between null \thetap\ photoproduction on protons \cite{g11APS}
vs. clear signal on neutrons (via deuteron target), 
e.g.~\cite{Kubarovsky:2003fi},\cite{Nakano2}.

A photon which turns into $K^+K^-$ can make a \thetap\,$K^-$
directly on a neutron, but cannot make a \thetap\ directly on a proton.
A photon which turns into $K^o \bar K^o$ can make a \thetap\,$\bar K^o$
directly on a proton, but cannot make a \thetap\ directly on a neutron.
So how much of the photon appears as $K^+K^-$ and how much as $K^o \bar K^o$
is an experimental question that can be clarified by measuring the neutral
kaons, e.g. comparing $\gamma p \rightarrow p K^+ K^-$ with
the analogous reaction with neutral kaons in the final state.
One can also do an analogous comparison  in $\gamma d$.
The relevant data might in principle be available in the CLAS g11 and 
LEPS experiments.

We also note that the $\gamma K^o \bar K^o$ vertex is
forbidden by an $SU(3)$ flavor selection rule and can be expected to be
suppressed in comparison  with the $\gamma K^+K^-$ vertex if $SU(3)_f$ is not
badly broken. 

The $\gamma K^o \bar K^o$ selection rule has been derived \cite{UspinGparity}
as an $SU(3)_f$  rotation of the $G$-parity \cite{RPP}
forbidden $\omega \rightarrow \pi^+ \pi^-$. 
This rotation turns isospin into the $U$-spin $SU(2)$  subgroup of
$SU(3)_f$ (analogous to isospin, interchanging $d \leftrightarrow s$)
\cite{Uspin}. The rotation turns  $\pi^+ \pi^-$   into $K^o \bar K^o$, 
while the photon is  a $U$-spin scalar \cite{UspinPhoton} particle, 
analogous to the isoscalar $\omega$.

In the vector dominance picture the photon is an $SU(3)_f$ octet and a $U$-spin
scalar combination of the  $\rho$, $\omega$ and $\phi$. In unbroken $SU(3)_f$
they are exactly degenerate and  their contributions to $\gamma \rightarrow K^o
\bar K^o$ cancel exactly.  In the real world $SU(3)_f$ breaking as measured by
vector meson masses is about 25\%-30\%, 
% $1- M(\rho/\omega) / M(\phi) \approx0.25$ 
so  the relative importance of the $\phi$ component is an open question. Further
implications of this breaking are discussed below. 

    Another simple way to see how the $\gamma K^o \bar K^o$ 
vertex vanishes in the $SU(3)$ flavor limit is by considering what happens
at the microscopic quark level.
The photon creates a single quark-antiquark
pair via a QED interaction. 
The other pair must be created via QCD by gluons. There are thus two
independent amplitudes in QED-QCD for creating this pair of composite
particles:
\begin{enumerate}
\item
The photon creates a $\bar d d$ pair and an $\bar s s$
 pair is added by gluon pair creation.
\item
The photon creates an $\bar s s$ pair and a $\bar d d$ 
pair is added by gluon pair creation.
\end{enumerate}
    Since the $d$ and $s$ quarks have the same electric charge and the same
couplings to gluons, these amplitudes are equal 
and add coherently in the $SU(3)$ flavor limit.
    However, if one writes down these two amplitudes carefully putting
in the momenta, one can see that if the first transition creates a $K^o$ with
momentum $p_1$ and a $\bar K^o$ with momentum $p_2$, 
the second transition creates a $K^o$
with momentum $p_2$ and a $\bar K^o$ with momentum $p_1$. 
The sum of these amplitudes therefore gives a state which is even under charge
conjugation.
    Since the photon is odd under charge conjugation, this $C$-even
amplitude cannot contribute to the final $C$-odd state and the transition
vanishes. 

This can be seen formally by writing down all the terms with
proper phases that reflect the photon being $C$-odd, i.e.  the matrix
element for a photon creating a quark with momentum $p_1$ and an antiquark
with momentum $p_2$ must be equal and opposite to the matrix element for a
photon creating a quark with momentum $p_2$ and an antiquark with momentum
$p_1$,
\beq{Cparity}
A\left[\gamma \rightarrow d(p_1) + \bar d(p_2)\right] 
=
{-} A\left[ \gamma \rightarrow d(p_2) + \bar d(p_1) \right]
\eeq
The photon is flavor-blind, so the amplitude on the r.h.s. is equal,
including sign, to the amplitude where $s$ quarks are replaced by $d$
quarks, so we have
\beq{CparityII}
A\left[\gamma \rightarrow d(p_1) + \bar d(p_2)\right] 
=
{-}
A\left[ \gamma \rightarrow s(p_2) + \bar s(p_1) \right]
\eeq
But when the primary $\bar q q$ pairs
 get ``dressed" by QCD, \eqref{CparityII} implies
\beq{CparityIII}
A\left[\gamma \rightarrow K^o(p_1) + \bar K^o(p_2)\right]
={-}
A\left[\gamma \rightarrow K^o(p_1) + \bar K^o(p_2)\right]
\eeq
which means that $\gamma K^o \bar K^o$ vertex must vanish.

    $SU(3)_f$ breaking by quark mass differences means that $d$ and $s$ quarks
with the same momenta have different energies. Thus the cancellation is no
longer exact.

The suppression of the $\gamma K^o \bar K^o$ vertex in the 
$\gamma N \rightarrow N K \bar K$ photoproduction reactions can be put on a
firmer foundation by using the unitarity relations for these reactions:
\beq{unitpNks}
\begin{array}{ccc}
{\rm Im}  \bra{K^+ K_s\, n} T \ket{\gamma p}  &=&
\ds
\kappa\sum_i  \bra{K^+ K_s\, n} T^{\dag} \ket {i} \bra{i}  T \ket{\gamma p} 
\hfill\\
\\
{\rm Im}  \bra{K^+ K^-n } T \ket{\gamma n}  &=&
\ds
\kappa\sum_i  \bra{K^+ K^-n} T^{\dag} \ket {i} \bra{i}  T \ket{\gamma n} 
\end{array}
\eeq
where $\kappa$ is a kinematic factor and the sum is over all intermediate
states $\ket {i}$. Intermediate states in the unitarity sum must be on the mass
shell,  and we  have seen that the photon coupling to  the $\gamma K^o \bar
K^o$ vertex is forbidden by an $SU(3)$ flavor selection rule and can be
expected to be suppressed in comparison  with the $\gamma K^+K^-$ vertex if
$SU(3)_f$ is not badly broken. We therefore assume that the unitarity sum is
dominated by the   $K^+K^- N$ intermediate state. Thus

\beq{unitpNks2}
\begin{array}{ccc}
{\rm Im}  \bra{K^+ K_s\, n} T \ket{\gamma p} &=&
\kappa \bra{K^+ K_s\, n} T^{\dag} \ket {K^+K^-p} \bra{K^+K^-p} T \ket{\gamma p} 
\\
\hfill\\
{\rm Im}  \bra{K^+ K^-n } T \ket{\gamma n}  &=&
\kappa \bra{K^+ K^-n} T^{\dag} \ket {K^+K^-n} \bra{K^+K^-n}  T \ket{\gamma n} 
\end{array}
\end{equation}
We now see that the transition matrix  
$\bra{K^+ K_s n} T \ket{\gamma p}$ is
proportional to the transition matrix  
$\bra{K^+ K_s n} T^{\dag} \ket {K^+K^-p}$,
where the $\Lambda(1520)$ can appear as a resonance but the \thetap\ cannot.

Conversely, the transition matrix $\bra{K^+ K^-n } T \ket{\gamma n}$ is
proportional to the transition matrix 
$\bra{K^+ K^-n} T^{\dag} \ket {K^+K^-n}$,
where the $\Lambda(1520)$ cannot appear as a resonance but the \thetap\ can.

This unitarity relation holds only at low enough energies so that transitions
to final states containing more particles can be neglected. The restriction to
intermediate states on mass shell eliminates the need to the include diagrams
with higher states found in many theoretical
treatments \cite{atsushi, close, Nam}. 
Note that the necessity to consider $K^*$ exchanges which arises in all such
theoretical treatments and in all treatments well above the $K^*$ production
threshold does not arise here.

Thus the g11 reaction $\gamma p \rightarrow n K^+ K_s$
proceeds via 
\,$\gamma p \rightarrow K^+K^- p \rightarrow K^+ K_s n$\,
and there is no possibility of making the $\Theta^+$
in any simple way in the g11 setup.

 The basic physics here is that the photon couples much more strongly
to charged kaons than to neutral kaons and charged kaons can make the
$\Theta^+$ simply on a neutron and not on a proton.

One test of this picture is to compare the photoproduction of isoscalar
baryon resonances with positive and negative strangeness 
on proton and neutron targets.

\beq{final}
\gamma p \rightarrow \bar K^o \Theta^+; ~ ~ ~ \gamma n \rightarrow K^- \Theta^+  
\end{equation}

\beq{Lambda}
\gamma n \rightarrow K^o \Lambda; ~ ~ ~ \gamma p \rightarrow K^+ \Lambda 
\end{equation}

Positive strangeness resonances like the \thetap\ will be produced on neutron
targets and not on protons, while negative strangeness resonances like the 
$\Lambda(1520)$ will be produced on proton 
targets and not on neutrons.

Recent data on $\Lambda(1520)$ production confirm this picture. 
LEPS observes a strong asymmetry in photoproduction of $\Lambda(1520)$ on 
proton and neutron \cite{Nakano2,Nam}.
They measured both \,
$\gamma p \rightarrow \Lambda(1520) K^+$ 
and 
$\gamma d \rightarrow \Lambda(1520) K N$ 
and find that the production rate on the deuteron is almost equal to and
even slightly smaller than on the proton. This implies that 
\hbox{$\gamma n \rightarrow \Lambda(1520) K_s$} is negligible.

We therefore propose the suppression of the $\gamma K^o \bar K^o$ vertex
as the likely explanation of 
non-observation of \thetap\ production on a proton target in the 
CLAS g11 experiment \cite{g11APS}. To gain confidence in this explanation 
it is important that other experiments confirm the
relative suppression of $\gamma n \rightarrow \Lambda(1520) K$.

\subsection{SU(3) breaking and the $\phi$ component of the photon}

In the vector dominance picture with broken $SU(3)_f$ and octet-singlet mixing
the  $\rho$ and $\omega$ components of the photon are still degenerate  but the
$\phi$ is now separate. In this picture for \thetap\ photoproduction the $\bar
s$ strange antiquark is already present in the initial state in the isoscalar
$\phi$ component. The $\rho$ and $\omega$ components contain no strangeness and
can produce the \thetap\  only via the production of an $s \bar s$ pair from 
QCD gluons.  How much this extra strangeness production costs is still open.
This cost does not appear in treatments using the Kroll-Ruderman theorem
\cite{Hosaka}  which involves only pions and ignores the $\phi$ component of
the photon.

One example of an experiment that projects out the $I=0$ state of a $KN$ state
is the photoproduction on a deuteron \cite{Nakano2}  of the final state 
$\Lambda(1520) K^+ n$.

If the strangeness in the reaction comes from the isoscalar $\phi$
component of the photon, the final $KN$ state is required by isospin invariance
to be isoscalar, and the signal is observed against a purely isoscalar 
background.  
This is not true for the other $K^-pK^+n$ final states  observed
in the same experiment where the  $K^-p$ is not in the $\Lambda(1520)$ and the
effects of a nonresonant $I=1$ background can give very different 
results.\footnote{We recall old SLAC experiments
\cite{Boyarski:1970yc, Quinn:1975} which looked at photoproduction of $K^+$-hyperon from
hydrogen and deuterium at 11 and 16 GeV. It would be interesting to
re-examine these data in view of \cite{Nakano2}.}

\section{Interference between resonances and background}

\subsection{Specific coherence effects in some experiments}
We also point out here one additional crucial factor which 
can explain why some
experiments see the \thetap\ and others do not. All experiments search for a
narrow resonance against a nonresonant background, generally with an
experimental resolution much coarser than the assumed \thetap\ width. Some
experiments lead to a definite final state, e.g the reactions 
(\ref{final}).

The probability of
observing the signal in such an experiment is very sensitive to the relative
phase between the signal and the background. This is shown explicitly below in
a toy model. 
Such interference effects are not expected in inclusive \thetap\ production;
e.g.
\beq{incl}
e^+e^- \rightarrow \Theta^+ X  
\end{equation}
Here the incoherent sum over all inclusive final states $X$ destroys all phase
information.

The \thetap\ is believed to decay into a kaon and nucleon with isospin zero.
But the observed decay modes \,$K^+n$ \,and\, $K^op$\, are equal mixtures of states with
$I=0$ and $I=1$ with opposite relative phase. 
If the nonresonant background in a 
given experiment is mainly from an $I=1$ amplitude, the relative phase between 
signal and background amplitudes will depend upon the particular decay mode
observed. The phase when
the \thetap\ is detected in the $K^+n$ decay mode will be opposite to the phase
in the same experiment where the \thetap\ is detected in the $K^op$ decay mode.

A serious isospin analysis may be necessary to understand the implications of
any experiment where destructive interference between an $I=0$ signal and 
$I=1$ background can effectively destroy the signal. Simple isospin relations
between similar reactions which consider only the signal and not the
interference with the background can give very erroneous results. In particular
one can expect apparent contradictions between negative and positive results
which are connected by isospin when the interference with background is
neglected.   How this destruction can occur is illustrated in the toy model
below.

\subsection{A simple model for resonance and background}

Consider a simple toy model for a resonance and background and write the 
amplitude for the production of this resonance as
 
\beq{resback}
A = b\cdot e^{i\phi} + {{1}\over{1+ix}}
\end{equation}
where $x$ denotes the difference between the energy and the resonance energy, 
the strength of the resonance amplitude is normalized to unity and 
$b$ and $\phi$ denote the amplitude and phase of the nonresonant background.
The background is assumed to be essentially constant over an energy region 
comparable to the width of a narrow resonance. 
The square of this amplitude is then

\beq{resback2}
\begin{array}{lcl}
|A|^2 = \ds \left| b\cdot e^{i\phi} + {{1}\over{1+ix}}\right|^2 
&=& 
\ds
b^2 +2 {\rm Re} \left({{b\cdot e^{i\phi}}\over{1-ix}}\right)+ 
{{1 }\over{1+ x^2}}
\\
\hfill\\
&=&
\ds
b^2 + {{1+2 b \cos \phi - 2 b x\sin \phi }\over{1+ x^2}} 
\end{array}
\end{equation}

For a detector which integrates the cross section over a symmetric
interval from $-X$ to $+X$ the term linear in $x$ drops out and
\beq{resback6}
\begin{array}{lcl}
\ds \int_{-X}^X |A|^2 dx &=& \ds \int_{-X}^X  dx \left(b^2 +
{{1+2 b \cos \phi}\over{1+ x^2}}\right) 
 = 2 b^2 X + \int_{-\Theta}^\Theta d \theta (1+2 b \cos \phi)  
\\
\hfill
\\
&=& 2 b^2 X + 2(1+2 b \cos \phi)\Theta
\end{array}
\end{equation}
where we have set $x = \tan\theta$ and $X=\tan\Theta$.

The ratio of signal to background is then
\beq{resback67}
{{\rm{signal}}\over{\rm{background}}} = 
{{1+2 b \cos \phi}\over{b^2}}\cdot {{\Theta}\over{\tan \Theta}}
\end{equation}

\begin{figure}[t]
\strut\vskip-1cm
\strut\kern1.5em
\epsfig{figure=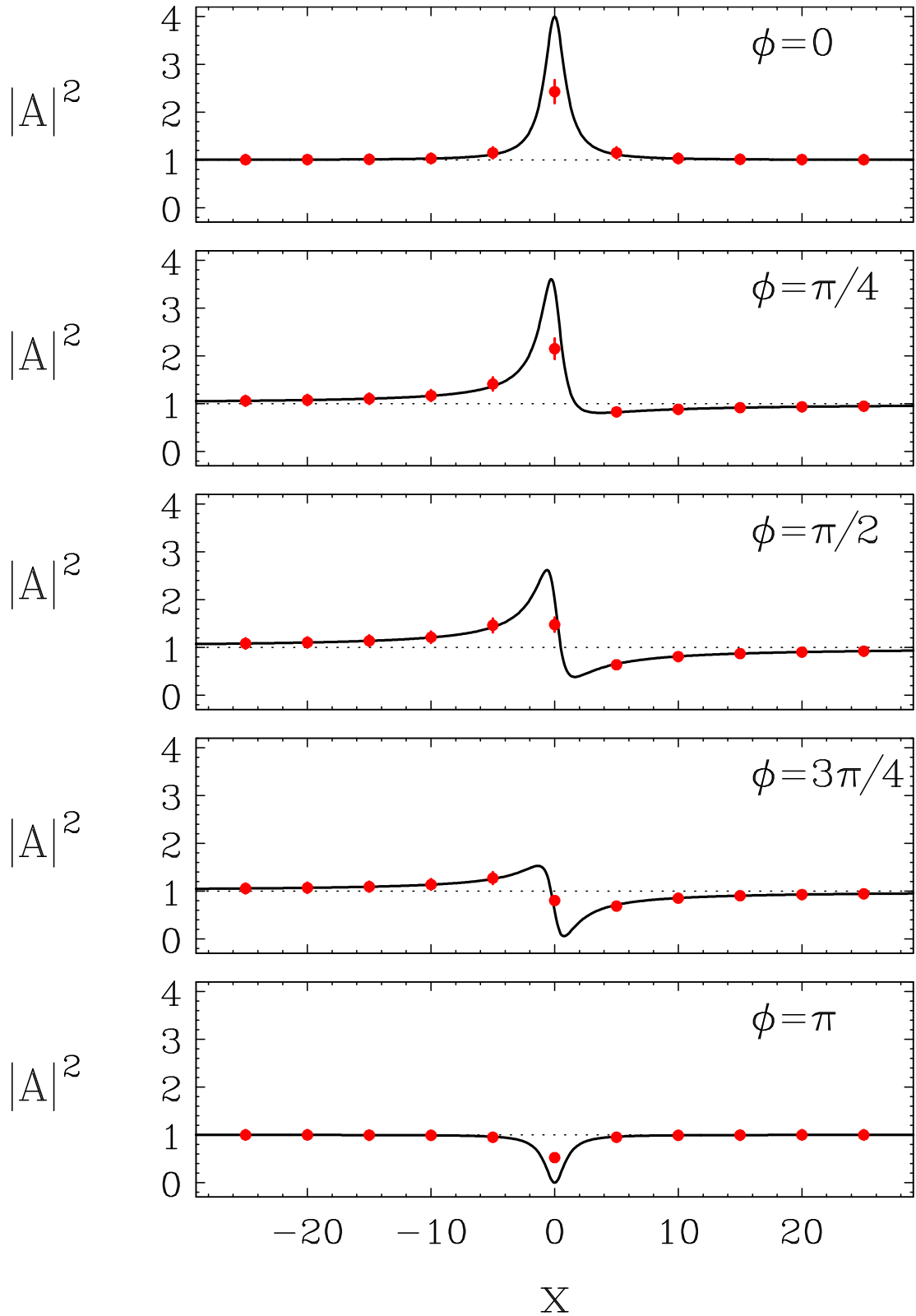,angle=90,width=0.90\textwidth}
\strut\vskip0cm
\caption{\it $|A|^2$, eq.~\eqref{resback2}, as function of $x$ for $b=1$ and
for different values of the relative phase,
$\phi = 0,\pi/4,\pi/2,3\pi/4$ and $\pi$. Red points denote values averaged
over bins of width $\Delta x=5$, assuming an additional 10\% relative
error.}
 \label{figI}
\end{figure}

We thus find that for the case where $b=1$, i.e. the signal and background
amplitudes are equal at the peak of the resonance, the ratio of integrated
signal to  integrated background varies from ${-}1$ to 3, depending upon the
relative phase.\footnote{Signal strength is defined as the difference
between the measured (signal+background) and background alone. Relative
signal strength of ${-}1$ corresponds to a dip, cf. $\phi=\pi$ entry
in Fig.~\ref{figI}.}

Fig.~\ref{figI} shows $|A|^2$ as function of $x$ for several
representative values of of the relative phase: 
$\phi = \ds 0,{\pi\over 4},{\pi\over2},{3\pi\over4}$ and $\pi$.

For illustration purposes we have taken $b=1$, i.e. equal strength of the
background and signal amplitudes at the peak of the resonance. To expound
the effect experimental resolution being substantially coarser than the
resonance width,  $|A|^2$ was averaged over bins of width $\Delta x =5$,
i.e. five times wider than the resonance width. The resulting values are
plotted at bin centres as red points, with additional 10\% relative error bars.
\clearpage % to force figure to be printed on next available page

The interplay of the limited experimental resolution with the relative
phase of signal and background can lead to rather striking results,
depending on the value of $\phi$. Thus e.g. for $\phi = \ds {3\pi\over4}$
the measured signal is almost completely washed out by this `conspiracy'
of the relative phase and the experimental resolution. 

\subsection{Mass difference in $ K^+ n $ and $K_s p$ modes}
In experiments which did observe the
\thetap\
there seems to be a small but systematic and non-negligible mass difference
between the mass observed in the
$ K^+ n $ and $K_s p$ modes  \cite{Zhao:2004rg}.

It is interesting to ask if this could have
something to do with isospin effects in signal-background interference.
As already mentioned,
$K^+n$ and $K_s p$ are equal mixtures of states with
$I=0$ and $I=1$ with opposite relative phase. 
If the nonresonant background in a
given experiment is mainly from an $I=1$ amplitude, the relative phase
between
signal and background amplitudes will depend upon the particular decay mode
observed. The phase when
the \thetap\ is detected in the $K^+n$ decay mode will be opposite to the
phase
in the same experiment where the \thetap\ is detected in the $K^op$ decay
mode. 

We now consider the effect of this phase flip in our simple model of
signal-background interference. A sign change of the relative phase 
in eq.~\eqref{resback} results in
\beq{PhaseFlip}
\phi \longrightarrow \phi + \pi;\qquad
A \longrightarrow 
\tilde A = b\cdot e^{i{\phi+\pi}} + {{1}\over{1+ix}}
\end{equation}
Since the observed signal is given by the absolute value of the amplitude,
we can replace $A$ by its complex conjugate, i.e.
\beq{PhaseFlip2}
\phi \longrightarrow \phi + \pi;\qquad
|A|^2 \rightarrow |A^*|^2= \left| b\cdot e^{i(\pi-\phi)} 
+ {{1}\over{1-ix}}\right|^2
\end{equation}
So if Fig.~\ref{figI} were to describe the $K^+ n$ mode, the corresponding
plots of $|A|^2$ for the $K_s p$ mode can be obtained by the transformation
$\phi \rightarrow \phi - \pi$, \ $x \rightarrow {-}x$, possibly resulting
in the shift in the peak location.

On the other hand, backgrounds differ from experiment to experiment, so it
is not at all clear why this would result in a systematic shift between 
$K^+ n$ and $K_s p$ modes. It would be very interesting if our experimental
colleagues could redo this
analysis with a realistic background parametrization.

\section{Conclusions}

The theory needed to understand apparent contradictions in considering the
question of why some experiments see the \thetap\ and others do not turns out
to be more complicated than  naively expected.

Before conclusions can be drawn from apparent violations
of isospin symmetry or from negative search results using a particular final
state extensive data and serious isospin analysis are needed.  In particular,

\begin{enumerate}
\item Data from both proton and neutron targets 
\item Data from both the neutral and charged kaon final states
\item Extensive isospin analyses that include the background 
\item Clarification of the role of the $\phi$ component of the photon  
\item Clarification of the degree suppression of
$\gamma \rightarrow K^o\bar K^o$ 
\end{enumerate}

Serious signal-background interference effects can occur in some experiments
which search for the \thetap\ as a narrow $I=0$ resonance  against  a
nonresonant background with an experimental resolution larger than the expected
resonance width. The resonance signal can be completely destroyed by
destructive  interference with a background having a magnitude of the same
order as the signal and a destructive phase.

Experiments detect \thetap\ via the decay modes  $K^+n$ and $K^op$ which are
equal mixtures of $I=0$ and $I=1$ states with opposite relative phase. Unless 
the experiment projects out the $I=0$ component of the signal, interference can
occur between the $I=0$ state and an $I=1$ background. Such interference can
vary greatly between final states which are related by isospin in the absence
of interference.

All these considerations must be understood before reliable conclusions can be
drawn.  But when some experiments do not see the \thetap, there are 
researchers who tend to ignore 
these questions and immediately conclude that the \thetap\ does not exist. 

In the meantime, we note that LEPS provides fresh evidence for 
photoproduction of \thetap\ on deuteron \cite{Nakano2}
and CLAS is expected to release the first g10 deuteron data very soon
\cite{HicksPC}. If \thetap\ is seen in the latter as well, then
understanding the seeming discrepancy between the null result in 
g11 proton data and the deuteron experiments will be on top of the
theoretical agenda. 

We think that the 
suppression of $\gamma K^o \bar K^o$ vertex 
and signal/background interference effects discussed here are a first
promising step towards in this direction. We eagerly await
additional experimental data which can help in resolving this issue.

\section*{Acknowledgements}

The research of one of us (M.K.) was supported in part by a grant from the
Israel Science Foundation administered by the Israel
Academy of Sciences and Humanities.
We thank
Marco Battaglieri,
Frank Close,
Ken Hicks,
Atsushi Hosaka,
Uri Karshon,
Takashi Nakano,
Stepan Stepanyan,
Mark Strikman and
Raffaella De Vita
for discussions and
Maxim Polyakov for a comment about a relation between 
coupling of the photon to neutral kaons and gauge invariance.

%----------------------------------------------------------------------
% This prevents REFERENCES from forcing a page break
%\def\newpage{\vskip10ex}
%
\catcode`\@=11 % This allows us to modify PLAIN macros
\def\references{
\ifpreprintsty \vskip 10ex
%\ifpreprintsty \newpage
%
\hbox to\hsize{\hss \large \refname \hss }\else
\vskip 24pt \hrule width\hsize \relax \vskip 1.6cm \fi \list
{\@biblabel {\arabic {enumiv}}}
{\labelwidth \WidestRefLabelThusFar \labelsep 4pt \leftmargin \labelwidth
\advance \leftmargin \labelsep \ifdim \baselinestretch pt>1 pt
\parsep 4pt\relax \else \parsep 0pt\relax \fi \itemsep \parsep \usecounter
{enumiv}\let \p@enumiv \@empty \def \theenumiv {\arabic {enumiv}}}
\let \newblock \relax \sloppy
 \clubpenalty 4000\widowpenalty 4000 \sfcode `\.=1000\relax \ifpreprintsty
\else \small \fi}
\catcode`\@=12 % at signs are no longer letters
%-----------------------------------------------------------------

\end{document}